\documentstyle[twoside,fleqn,espcrc2,epsf]{article}

\makeatletter
\long\def\@makefigurecaption#1#2{\vskip 0mm #1. #2\par}
\makeatother
\def\Pacc{P_{\hbox{\tiny acc}}}     
\def\dH{\delta H}                   
\def\dt{\delta\tau}                 
\def\scrm{{\cal M}}                 
\def\tr{\mathop{\rm tr}}            

\title{A New Exact Method for Dynamical Fermion Computations with
  Non-Local Actions}

\author{Ivan Horv\'ath\address{Department of Physics,
        University of Virginia,
        382 McCormick Road,
        Charlottesville, VA~22903, USA}
        and
        A. D. Kennedy
        \address{SCRI, Florida State University,
        Tallahassee, FL~32306, USA}
        \thanks{Current address: Maxwell Institute,
        King's Buildings, The University of Edinburgh,
        Mayfield Road, Edinburgh EH9 3JZ, Scotland}
        \thanks{Speaker}
        and
        Stefan Sint${}^{\rm b}$
        \thanks{Current address: Dipartimento di Fisica,
        Universita degli Studi di Roma Tor Vergata,
        Via della Ricerca Scientifica~1,
        I--00133 Rome, Italy}}
\begin{document}

\begin{abstract}
  We introduce a new algorithm which we call the {\em Rational Hybrid
  Monte Carlo} Algorithm (RHMC). This method uses a rational
  approximation to the fermionic kernel together with a noisy
  (Kennedy--Kuti~\cite{kennedy85e}) acceptance step to give an
  efficient algorithm with no molecular dynamics integration step-size
  errors.
\end{abstract}

\maketitle

\section{Introduction}

The RHMC algorithm is an exact method for generating lattice configurations
distributed according to some non-local action. Both the terms `non-local' and
`exact' require some explanation.

By a non-local action we mean one which is constructed from some continuous
function of a local kernel, for example $\bar\psi\sqrt\scrm\psi$ where $\scrm$
is the usual staggered kernel; we do not mean an action with complicated
long-range interactions and many coupling constants such as a `perfect'
action. Two examples of interesting non-local actions are two flavours of
staggered fermions, and solutions of the Ginsparg--Wilson relation such as
Neuberger's action~\cite{Neuberger:1997fp,Luscher:1998pq,Hasenfratz:1998ri}.%
\footnote{How and if our methods can be applied to Neuberger's action is
currently under investigation.}

Our goal is not to consider the virtues or otherwise of such actions, but only
to discuss how they might be efficiently simulated.

`Exact' algorithms in our sense are ones which do not require a zero
integration step-size extrapolation. This is important in the case of two
flavours of staggered fermions in order to determine the order of phase
transitions, for example.

\section{Ideas underlying the algorithm}

Our method is a Hybrid Monte Carlo (HMC) algorithm with two new ingredients,
a cheap accept/reject step, and a cheap force computation.

For the accept/reject step, which is needed to make the algorithm exact, the
problem is that it is too expensive to compute the change in energy $\dH$
exactly. We therefore apply the Kennedy--Kuti method~\cite{kennedy85e} and use
a {\em linear} accept/reject step by defining an ordering of the fields
independent of the estimator of the action. The number of violations of
$0\leq\Pacc\leq1$ can be counted explicitly and easily be made negligibly
small by a suitable choice of algorithmic parameters. In order to produce an
unbiased stochastic estimate of $\exp\tr\ln\scrm$ we sum the series expansions
for $e^x$ and $\ln(1-x)$ stochastically~\cite{kennedy85f}.

\subsection{Noisy Force} The computation of the force needed for the molecular
dynamics (MD) evolution is also prohibitively expensive for non-local actions,
and we first investigated modifying the Hybrid Molecular Dynamics (HMD) $R_0$
algorithm by adding an accept/reject step to make it exact. This did not lead
to a feasible algorithm because the integration errors grow as $\dH\propto
V\dt$, and thus the cost grows as $V(V\dt)(\xi/\dt) = V^2\xi$ where $\dt$ is
the integration step size, $\xi$ is the correlation length in units of MD time
and $V$ is the lattice volume. This cannot be improved using higher-order
integration schemes because the errors are intrinsically due to the noise. The
$R$ algorithm \cite{gottlieb87a} is not area-preserving or reversible so cannot
be made exact in any obvious fashion.

\subsection{Polynomial Force} Instead of using a noisy force we then observed
that the HMC algorithm does not require that we integrate the classical
equations of motion for the Hamiltonian corresponding to the action we are
interested in: any area-preserving reversible mapping on phase space suffices
to give a valid algorithm. Using an MD Hamiltonian which approximates the
desired one sufficiently well to give a good acceptance rate is a good choice,
and following L\"uscher~\cite{luescher94a} we can do this using a polynomial
in $\scrm$ with a minimax error over the compact spectrum of $\scrm$ which
falls exponentially with the degree of the polynomial.

The existence of such optimal polynomials was shown by Chebyshev, and a
truncated expansion in Chebyshev polynomials is usually very good, although in
general not optimal. The optimal polynomials can be computed easily using the
Remez algorithm. It is not obvious that the minimax norm is the best for our
purposes, indeed Montvay uses an $L^2$ norm instead, but it seems to be a safe
and adequate choice.

Such polynomial actions have been used before
\cite{jansen97a,forcrand96a,montvay95a}, but not in combination with a noisy
acceptance step for non-local actions. The major difficulty with using such
high-degree polynomial actions is that they can be very sensitive to rounding
errors in numerical computation, and much work has gone into choosing
orderings of the roots of the polynomial which minimise these
difficulties~\cite{Bunk:1998rm}.

\subsection{Rational Force} 
\begin{figure}[t]
  \begin{center}
    \epsfxsize=0.45\textwidth
    \epsfbox{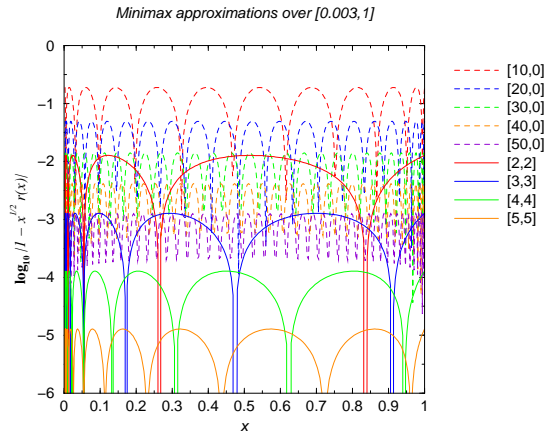}
  \end{center}
  \caption{Comparison of minimax errors for rational functions of degree
    $[n,d]$. The dashed curves (with $d=0$) are polynomials.}
  \label{fig:minimax-errors}
\end{figure}

It is well-known in the numerical analysis
literature that rational function approximations not only share the exponential
convergence property of polynomial approximations, but also usually reach a
sufficiently small minimax error for much lower degree.  The existence of
optimal rational approximations is easily established using Chebyshev's
arguments, but in this case we must use an iterative method such as the Remez
algorithm to find the optimal approximation. If we choose to use a diagonal
rational approximation, that is one in which the numerator and denominator are
polynomials of equal degree, then it turns out that the roots of both are all
real and negative. An example of such an approximation is
\begin{eqnarray*}
  \lefteqn{\textstyle {1\over\sqrt x} \approx
    {\scriptstyle 0.3904603901} \times} \\
  && \!\!\!\!\!\!\!\textstyle\times
    {(x + 2.3475661045) (x + 0.1048344600) (x + 0.0073063814) \over
     (x + 0.4105999719) (x + 0.0286165446) (x + 0.0012779193)}.
\end{eqnarray*}
The evaluation of such a rational approximation for a matrix requires several
conjugate gradient inversions with increasing masses.\footnote{We are also
investigating using a partial fraction expansion of the rational function and
using a multiple mass solver.} Figure~\ref{fig:minimax-errors} illustrates the
quality of the minimax approximations.

\section{Numerical results}

\begin{figure}[t]
  \begin{center}
    \epsfxsize=0.45\textwidth
    \epsfbox{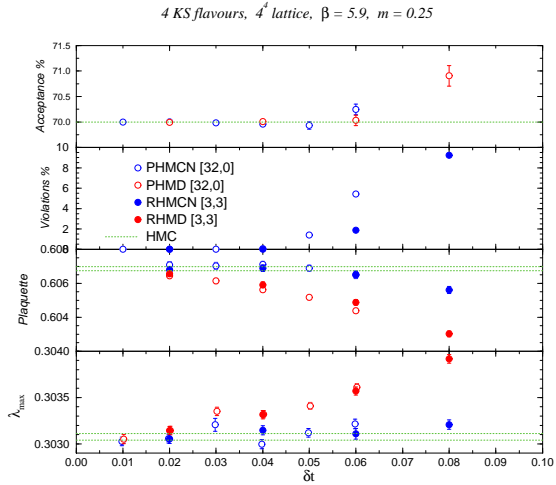}
  \end{center}
  \caption{Numerical Results for $N_f=4$.}
  \label{fig:nf=4}
\end{figure}

\begin{figure}[t]
  \begin{center}
    \epsfxsize=0.45\textwidth
    \epsfbox{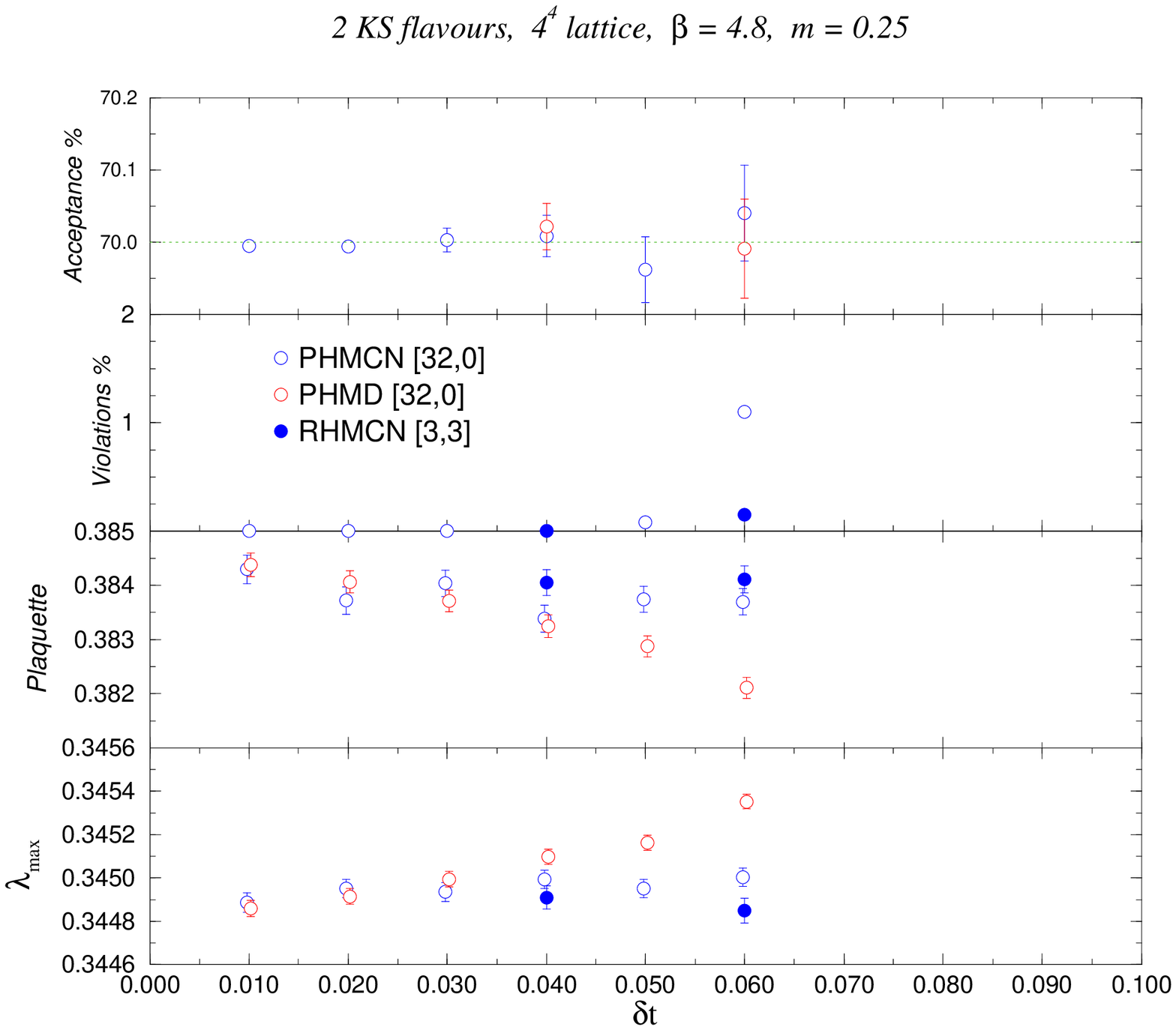}
  \end{center}
  \caption{Numerical Results for $N_f=2$.}
  \label{fig:nf=2}
\end{figure}

The results of our numerical tests of the RHMC algorithm are shown in
Figure~\ref{fig:nf=4} for the case of four flavours of staggered fermions,
where the results may be compared with a conventional HMC computation, and in
Figure~\ref{fig:nf=2} for the case of two staggered flavours, where there is no
other exact method available for comparison.

Each of these figures is composed of four graphs: the top one shows the mean
acceptance rate, which for our choice of parameters should be exactly 70\%; the
second one shows the number of violations, when this is zero the method is
exact; the third one shows the mean plaquette; and the bottom one the largest
eigenvalue of the staggered Dirac operator. The solid circles are results using
a $[3,3]$ rational approximation for the force and the open ones are for a
$[32,0]$ polynomial approximation.  The lines are the results from a
conventional HMC computation when this was possible.

Results are shown both for the algorithm with the noisy accept/reject step
(indicated by the suffix N) and for an inexact version without the
accept/reject step (indicated by the suffix D). The $\dt^2$ step size errors in
the latter are evident, as are the difficulties in carrying out an accurate
zero step size extrapolation.

\def\href#1{}
\def\ptitle#1{}

\end{document}